\def\year{2019}\relax
\documentclass[letterpaper]{article}  
\usepackage{aaai19}  			
\usepackage{times}  				
\usepackage{helvet}  			
\usepackage{courier}  			
\usepackage{url}  				
\usepackage{graphicx}  			
\usepackage[font=footnotesize,labelfont={bf},labelsep=colon]{caption}

\newcommand{\citet}[1]{\citeauthor{#1}~(\citeyear{#1})}

\usepackage{booktabs}
\usepackage{tabularx}
\usepackage{amsmath,amssymb,amsfonts}
\usepackage{multirow}
\usepackage{color}
\begin{document}

\title{Higher-order Graph Convolutional Networks}

\author{John Boaz Lee\textsuperscript{1}, Ryan A. Rossi\textsuperscript{2}, Xiangnan Kong\textsuperscript{1}, Sungchul Kim\textsuperscript{2}, Eunyee Koh\textsuperscript{2}, and Anup Rao\textsuperscript{2}\\
\textsuperscript{1}{Worcester Polytechnic Institute, MA, USA}\\
\textsuperscript{2}{Adobe Research, CA, USA}\\
\{jtlee, xkong\}\!@wpi.edu, \{rrossi, sukim, eunyee, anuprao\}\!@adobe.com
}
\maketitle

\begin{abstract}
Following the success of deep convolutional networks in various vision and speech related tasks, researchers have started investigating generalizations of the well-known technique for graph-structured data. A recently-proposed method called Graph Convolutional Networks has been able to achieve state-of-the-art results in the task of node classification. However, since the proposed method relies on localized first-order approximations of spectral graph convolutions, it is unable to capture higher-order interactions between nodes in the graph. In this work, we propose a motif-based graph attention model, called Motif Convolutional Networks, which generalizes past approaches by using weighted multi-hop motif adjacency matrices to capture higher-order neighborhoods. A novel attention mechanism is used to allow each individual node to select the most relevant neighborhood to apply its filter. Experiments show that our proposed method is able to achieve state-of-the-art results on the semi-supervised node classification task. 
\end{abstract}

\section{Introduction}
In recent years, deep learning has made a significant impact on the field of computer vision. Various deep learning models have achieved state-of-the-art results on a number of vision-related benchmarks. In most cases, the preferred architecture is a Convolutional Neural Network (CNN). CNN models have been applied successfully to the tasks of image classification~\cite{ImageNet}, image super-resolution~\cite{DRCN}, and video action recognition~\cite{action-ConvNets}, among many others.

CNNs, however, are designed to work for data that can be represented as grids (\textit{e.g.}, videos, images, or audio clips) and do not generalize to graphs -- which have more irregular structure. Due to this limitation, it cannot be applied directly to many real-world problems whose data come in the form of graphs -- social networks~\cite{deepwalk} or citation networks~\cite{ICA} in social network analysis, for instance.

A recent deep learning architecture, called Graph Convolutional Networks (GCN)~\cite{GCN} approximates the spectral convolution operation on graphs by defining a layer-wise propagation that is based on the one-hop neighborhood of nodes. The first-order filters used by GCNs were found to be useful and have allowed the model to beat many established baselines in the semi-supervised node classification task.

\begin{figure}[t]
\centering
\includegraphics[width=0.85\linewidth]{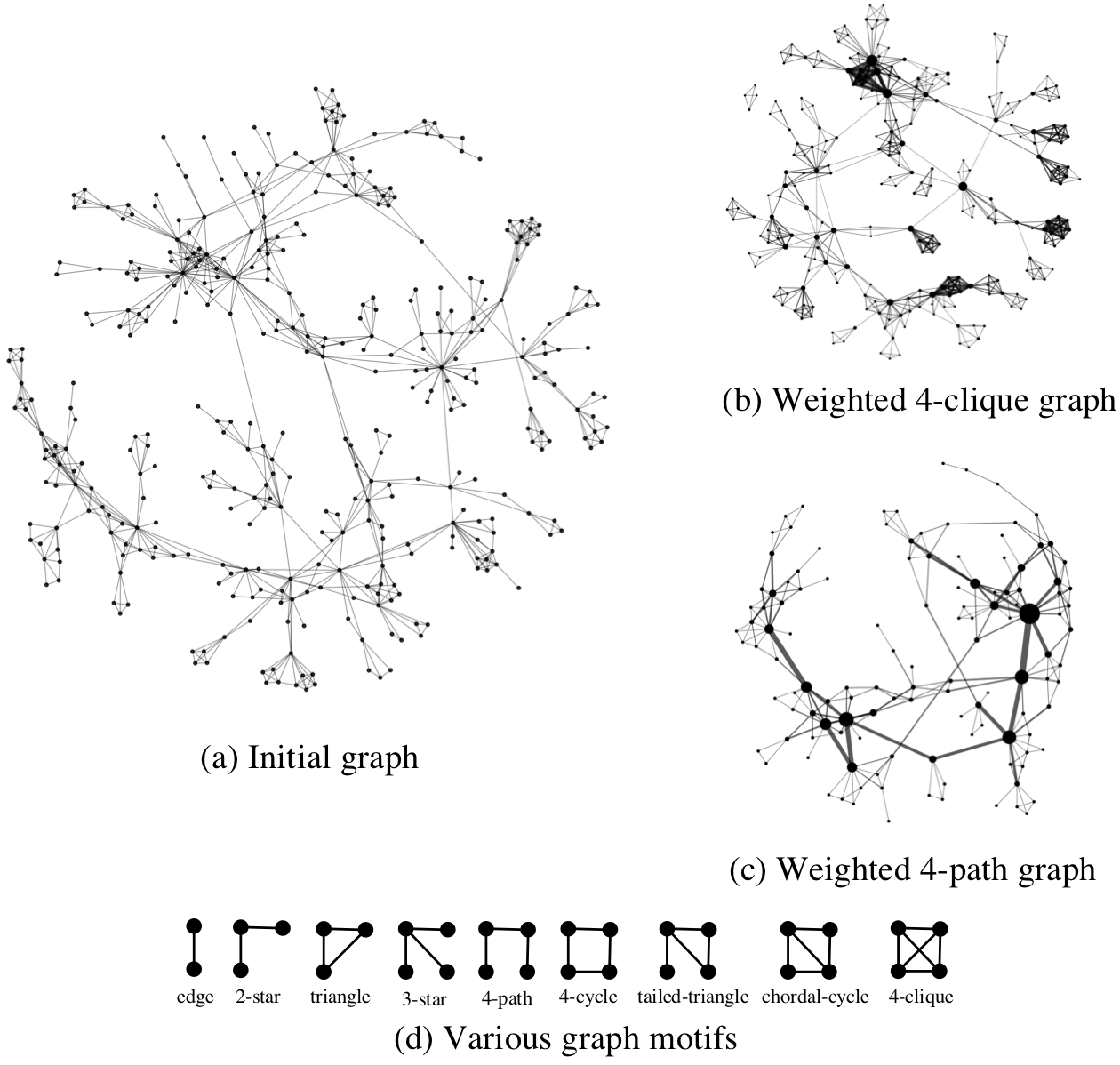}
\vspace{-3mm}
\caption{Graphs defined by different motif adjacencies can differ significantly. Size (weight) of nodes and edges in (b) and (c) correspond to the frequency of 4-node cliques and 4-node paths between nodes, respectively.}
\label{fig:butterfly}
\vspace{-5mm}
\end{figure}

However, in many cases, it has been shown that it may be beneficial to consider the higher-order structure in graphs~\cite{NEST,HONE,motifs-science,tnnls18-est}. In this work, we introduce a general class of graph convolution networks which utilize weighted multi-hop \emph{motif} adjacency matrices~\cite{HONE} to capture higher-order neighborhoods in the graph. 
The weighted adjacency matrices are computed using various network motifs~\cite{HONE}.
Fig.~\ref{fig:butterfly} shows an example of the node neighborhoods that are induced when we consider two different kinds of motifs, showing that the choice of motif can significantly alter the neighborhood structure of nodes.

Our proposed method, which we call Motif Convolutional Networks (MCN), uses a novel attention mechanism to allow each node to select the most relevant motif-induced neighborhood to integrate information from. Intuitively, this allows a node to select its one-hop neighborhood (as in classical GCN) when its immediate neighborhood contains enough information for the model to classify the node correctly but gives it the flexibility to select an alternative neighborhood (defined by higher-order structures) when the information in its immediate vicinity is too sparse and/or noisy.

The aforementioned attention mechanism is trained using reinforcement learning which rewards choices (\textit{i.e}, actions) that consistently result in a correct classification.
Our main contributions can be summarized as follows:

\begin{itemize}
	\item We propose a model that generalizes GCNs by introducing multiple weighted motif-induced adjacencies that capture various higher-order neighborhoods.
	\item We introduce a novel attention mechanism that allows the model to choose the best neighborhood to integrate information from.
	\item We demonstrate the superiority of the proposed method by comparing against strong baselines on three established graph benchmarks. We also show that it works comparatively well on graphs exhibiting heterophily.
	\item We demonstrate the usefulness of attention by showing how different nodes prioritize different neighborhoods.
\end{itemize}

\section{Related Literature}
\textbf{Neural Networks for Graphs } Initial attempts to adapt neural network models to work with graph-structured data started with recursive models that treated the data as directed acyclic graphs~\cite{graph-rnn2,graph-rnn1}. Later on, more generalized models called Graph Neural Networks (GNN) were introduced to process arbitrary graph-structured data~\cite{gnn1,gnn2}.

Recently, with the rise of deep learning and the success of models such as recursive neural networks (RNN)~\cite{rnn1,rnn2} for sequential data and CNNs for grid-shaped data, there has been a renewed interest in adapting some of these approaches to arbitrary graph-structured data.

Some work introduced architectures tailored for more specific problem domains~\cite{gated-gnn,NeuralFPS} -- like NeuralFPS~\cite{NeuralFPS} which is an end-to-end differentiable deep architecture which generalizes the well-known Weisfeiler-Lehman algorithm for molecular graphs -- while others defined graph convolutions based on spectral graph theory~\cite{spectral-gnn2}. Another group of methods attempt to substitute principled-yet-expensive graph convolutions using spectral approaches by using approximations of such.~\citet{cheby} used Chebyshev polynomials to approximate a smooth filter in the spectral domain while GCNs~\cite{GCN} further simplified the process by using first-order filters.

The model introduced by~\citet{GCN} has been shown to work well on a variety of graph-based tasks~\cite{gcn-task1,gcn-task2,GCN} and have spawned variants including~\cite{GAT,N-GCN}. We introduce a generalization of GCN~\cite{GCN} in this work but we differ from past approaches in two main points: first, we use weighted motif-induced adjacencies to expand the possible kinds of node neighborhoods available to nodes, and secondly, we introduce a novel attention mechanism that allows each node to select the most relevant neighborhood to diffuse (or integrate) information.

\noindent \textbf{Higher-order Structures with Network Motifs } 
Network motifs~\cite{motifs-science} are fundamental building blocks of complex networks; investigation of such patterns usually lead to the discovery of crucial information about the structure and the function of many complex systems that are represented as graphs.~\citet{motifs-bio} studied motifs in biological networks showing that the dynamical property of robustness to perturbations correlated highly to the appearance of certain motif patterns while~\citet{motifs-temporal} looked at motifs in temporal networks showing that graphs from different domains tend to exhibit very different organizational structures as evidenced by the type of motifs present.

Multiple work have demonstrated that it is useful to account for higher-order structures in different graph-based ML tasks~\cite{HONE,NEST,role2vec}.~DeepGL~\cite{deepGL} uses motifs as a basis to learn deep inductive relational functions that represent compositions of relational operators applied to a base graph function such as triangle counts.~\citet{HONE} proposed the notion of \emph{higher-order network embeddings} and demonstrated that one can learn better embeddings when various motif-based matrix formulations are considered.~\citet{NEST} defined a hierarchical motif convolution for the task of subgraph identification for graph classification. In contrast, we propose a new class of higher-order network embedding methods based on graph convolutions that uses a novel motif-based attention for the task of semi-supervised node classification.

\noindent \textbf{Attention Models } Attention was popularized in the deep learning community as a way for models to attend to important parts of the data~\cite{attention1,attention2}. The technique has been successfully adopted by models solving a variety of tasks. For instance, it was used by~\citet{attention1} to take glimpses of relevant parts of an input image for image classification; on the other hand,~\citet{attention3} used attention to focus on task-relevant parts of an image for the image captioning task. Meanwhile,~\citet{attention2} utilized attention for the task of machine translation by fixing the model attention on specific parts of the input when generating the corresponding output words. 

There has also been a surge in interest at applying attention to deep learning models for graphs. The work of~\citet{GAT} used a node self-attention mechanism to allow each node to focus on features in its neighborhood that were more relevant while~\citet{GAM} used attention to guide a walk in the graph to learn an embedding for the graph. More specialized methods of graph attention models include~\cite{GRAM,knowledge-attention} with~\citet{GRAM} using attention on a medical ontology graph for medical diagnosis and~\citet{knowledge-attention} using attention on a knowledge graph for the task of entity link prediction. Our approach differs significantly, however, from previous approach in that we use attention to allow our model to select task relevant neighborhoods.

\section{Approach}
We begin this section by introducing the foundational layer that is used to construct arbitrarily deep motif convolutional networks. When certain constraints are imposed on our model's architecture, the model degenerates into a GAT~\cite{GAT} which, in turn, generalizes a GCN~\cite{GCN}. Because of this, we briefly introduce a few necessary concepts from~\cite{GAT,GCN} before defining the actual neural architecture we employ -- including the reinforcement learning strategy we use to train our attention mechanism. 

\subsection{Graph Self-Attention Layer}
A multi-layer GCN~\cite{GCN} is constructed using the following layer-wise propagation:
\begin{align}
\label{eq:gcn}
\mathbf{H}^{(l+1)} = \sigma(\tilde{\mathbf{D}}^{-\frac{1}{2}} \tilde{\mathbf{A}} \tilde{\mathbf{D}}^{-\frac{1}{2}} \mathbf{H}^{(l)} \mathbf{W}^{(l)}).
\end{align}

\noindent Here, $\tilde{\mathbf{A}} = \mathbf{A} + \mathbf{I}_{N}$ is the modified adjacency matrix of the input graph with added self-loops -- $\mathbf{A}$ is the original adjacency matrix of the input undirected graph with $N$ nodes while $\mathbf{I}_{N}$ represents an identity matrix of size $N$. The matrix $\tilde{\mathbf{D}}$, on the other hand, is the diagonal degree matrix of $\tilde{\mathbf{A}}$ (\textit{i.e.}, $\tilde{D}_{i,i} = \sum_j \tilde{A}_{i,j}$). Finally, $\mathbf{H}^{(l)}$ is the matrix of node features inputted to layer $l$ while $\mathbf{W}^{(l)}$ is a trainable embedding matrix used to embed the given inputs (typically to a lower dimension) and $\sigma$ is a non-linearity.

The term $\tilde{\mathbf{D}}^{-\frac{1}{2}} \tilde{\mathbf{A}} \tilde{\mathbf{D}}^{-\frac{1}{2}}$ in Eq.~\ref{eq:gcn} produces a symmetric normalized matrix which update's each nodes representation via a weighted sum of the features in a node's one-hop neighborhood (the added self-loop allows the model to include the node's own features). Each link's strength (\textit{i.e.}, weight) is normalized by considering the degrees of the corresponding nodes. Formally, at each layer $l$, node $i$ integrates neighboring features to obtain a new feature/embedding via
\begin{align}
\label{eq:gcn-integrate}
\vec{\mathbf{h}}^{(l+1)}_{i} = \sigma\left( \sum_{j \in \mathcal{N}^{(\tilde{\mathbf{A}})}_{i}} \alpha_{i,j} \vec{\mathbf{h}}^{(l)}_{j} \mathbf{W}^{(l)} \right),
\end{align}

\noindent where $\vec{\mathbf{h}}^{(l)}_{i}$ is the feature of node $i$ at layer $l$, with fixed weights $\alpha_{i,j} = \frac{1}{\sqrt{\text{deg}(i) \, \text{deg}(j)}}$, and $\mathcal{N}^{(\tilde{\mathbf{A}})}_i$ is the set of $i$'s neighbors defined by the matrix $\tilde{\mathbf{A}}$ -- which includes itself.

In GAT~\cite{GAT}, Eq.~\ref{eq:gcn-integrate} is modified with weights $\alpha$ that are differentiable or trainable and this can be viewed as follows,
\begin{align}
\label{eq:gat-integrate}
\alpha_{i,j} = \frac{\text{exp} \left( \text{LeakyReLU} \left( \mathbf{a} [\vec{\mathbf{h}}_i \mathbf{W} \,\, \vec{\mathbf{h}}_j \mathbf{W} ]\right)\right)}{\sum_{k \in \mathcal{N}_i^{(\tilde{\mathbf{A}})}} \text{exp} \left( \text{LeakyReLU} \left( \mathbf{a} [\vec{\mathbf{h}}_i \mathbf{W} \,\, \vec{\mathbf{h}}_k \mathbf{W}]\right)\right)}.
\end{align}

\noindent The attention vector $\mathbf{a}$ in Eq.~\ref{eq:gat-integrate} is a trainable weight vector that assigns importance to the different neighbors of $i$ allowing the model to highlight particular neighboring node features that are more task-relevant.

Using the formulation in Eq.~\ref{eq:gat-integrate} with Eqs.~\ref{eq:gcn} and~\ref{eq:gcn-integrate}, we can now define multiple layers which can be stacked together to form a deep GCN (with self-attention) that is end-to-end differentiable. The initial input to the model can be set as $\mathbf{H}^{(1)} = \mathbf{X}$, where $\mathbf{X} \in \mathbb{R}^{N \times D}$ is the initial node attribute matrix with $D$ attributes. The final layer's weight matrix can also be set accordingly to output node embeddings at the desired output dimensions.

\subsection{Convolutional Layer with Motif Attention}
\begin{figure}[t]
\centering
\includegraphics[width=0.9\linewidth]{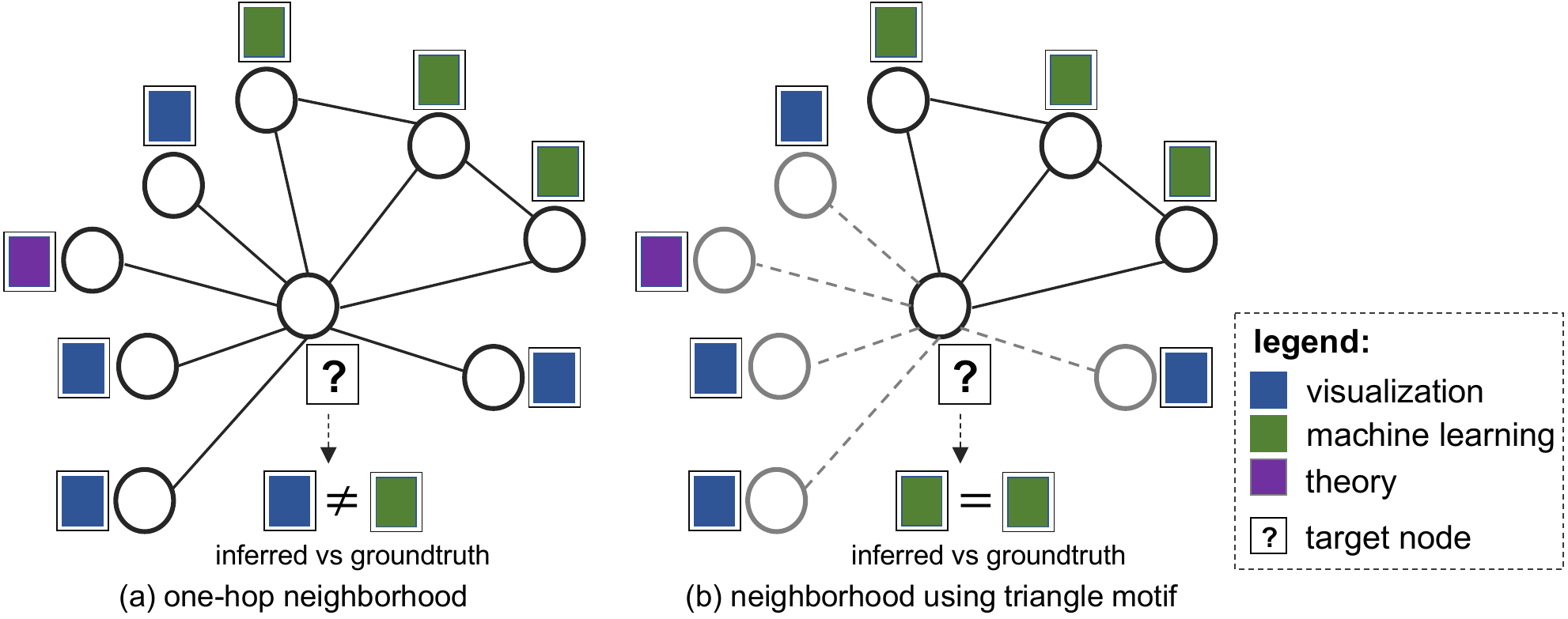}
\vspace{-2mm}
\caption{A researcher (target node) may have collaborated on various projects in visualization and theory. However, his main research focus is ML and he collaborates closely with lab members who also work among themselves. (a) If we simply use the target node's one-hop neighborhood, we may incorrectly infer his research area; however, (b) when we limit his neighborhood using the triangle motif, we reveal neighbors connected via stronger bonds giving us a better chance at inferring the correct research area. This observation is empirically shown in our experimental results.}
\label{fig:motiv}
\vspace{-5mm}
\end{figure}

We observe that both GCN and GAT rely on the one-hop neighborhood of nodes (\textit{i.e.}, $\mathbf{\tilde{\mathbf{A}}}$ in Eq.~\ref{eq:gcn}) to propagate information. However, it may not always be suitable to apply a single uniform definition of node neighborhood for all nodes. For instance, we show an example in Fig.~\ref{fig:motiv} where a node can benefit from using a neighborhood defined using triangle motifs to keep only neighbors connected via a stronger bond which is a well-known concept from social theory allowing us to distinguish between weaker ties and strong ones via the triadic closure~\cite{triadic-closure}.

\subsubsection{Weighted Motif-Induced Adjacencies}
Given a network $\mathcal{G} = (\mathcal{V}, \mathcal{E})$ with $N = |\mathcal{V}|$ nodes, $M = |\mathcal{E}|$ edges, as well as a set of $T$ network motifs\footnote{We use the term motifs loosely here and it can also be used to mean graphlets or orbits~\cite{graphlet}.} $\mathcal{H} = \{H_1, \cdots, H_T \}$, we can construct $T$ different motif-induced adjacency matrices $\mathcal{A} = \{\mathbf{A}_1, \cdots, \mathbf{A}_T \}$ with: $$(\mathbf{A}_t)_{i,j} = \text{\# of motifs of type } H_t \text{ which contain } (i,j) \in \mathcal{E}.$$
\noindent As shown in Fig.~\ref{fig:butterfly}, neighborhoods defined by different motifs can vary significantly. Furthermore, the weights in a motif-induced adjacency $\mathbf{A}_t$ can also vary as motifs can appear in varying degrees of frequency between different pairs of nodes. 

\subsubsection{Motif Matrix Functions} Each of the calculated motif adjacencies $\mathbf{A}_t \in \mathcal{A}$ can now be potentially used to define motif-induced neighborhoods $\mathcal{N}^{(\mathbf{A}_t)}_i$ with respect to a node $i$. While Eq.~\ref{eq:gat-integrate} defines self-attention weights over a node's neighborhood, the initial weights in $\mathbf{A}_t$ can still be used as reasonable initial estimates of each neighbor's ``importance.''

Hence, we introduce a \textit{motif-based matrix formulation} as a function $\Psi : \mathbb{R}^{N \times N} \rightarrow \mathbb{R}^{N \times N}$ over a motif adjacency $\mathbf{A}_t \in \mathcal{A}$ similar to~\cite{HONE}. Given a function $\Psi$, we can obtain \textit{motif-based matrices} $\tilde{\mathbf{A}}_t = \Psi(\mathbf{A}_t)$, for $t = 1, \cdots, T$. Below, we summarize the different variants of $\Psi$ that we chose to investigate.

\textbf{$\bullet$ Unweighted Motif Adjacency w/ Self-loops: } In the simplest case, we can construct $\tilde{\mathbf{A}}$ (here on, we omit the subscripts $t$ for brevity) from $\mathbf{A}$ by simply ignoring the weights:
\begin{equation}
\label{eq:k1}
\tilde{\mathbf{A}}_{i,j} =
\begin{cases} 
      1 & i = j \\
      1 & \mathbf{A}_{i,j} > 0 \\
      0 & \text{otherwise.}
   \end{cases}
\end{equation}

\noindent But, as mentioned above, we lose the initial benefit of leveraging the weights in the motif-induced adjacency $\mathbf{A}$. 

\textbf{$\bullet$ Weighted Motif Adjacency w/ Row-wise Max: } We can also choose to retain the weighted motif adjacency $\mathbf{A}$ without modification save for added row-wise maximum self-loops. This is defined as
\begin{align}
\label{eq:k2}
\tilde{\mathbf{A}} = \mathbf{A} + \mathbf{M},
\end{align}
\noindent where $\mathbf{M}$ is a diagonal square matrix with $M_{i,i} = \text{max}_{1\leq j \leq N} A_{i,j}$. Intuitively, this allows us to assign an equal amount of importance to a self-loop consistent with that given to each node's most important neighbor.

\textbf{$\bullet$ Motif Transition w/ Row-wise Max: }The random walk on the weighted graph with added row-wise maximum self-loops has transition probabilities $P_{i,j} = \frac{A_{i,j}}{(\sum_{k} A_{i,k}) + (\text{max}_{1 \leq k \leq N} A_{i,k})}$. Our random walk motif transition matrix can thus be calculated by 
\begin{align}
\label{eq:k3}
\tilde{\mathbf{A}} = \mathbf{D}^{-1} (\mathbf{A} + \mathbf{M}),
\end{align}
\noindent where, in this context, the matrix $\mathbf{D}$ is the diagonal square degree matrix of $\mathbf{A} + \mathbf{M}$ (\textit{i.e.}, $D_{i,i} = (\sum_{k} A_{i,k}) + (\text{max}_{1 \leq k \leq N} A_{i,k})$) while $\mathbf{M}$ is defined as above. Here, $\tilde{A}_{i,j} = P_{i,j}$ or the transition probability from node $i$ to $j$ is proportional to the motif count between nodes $i$ and $j$ relative to the total motif count between $i$ and all its neighbors. 

\textbf{$\bullet$ Absolute Motif Laplacian: }The absolute Laplacian matrix can be constructed as follows:
\begin{align}
\label{eq:k4}
\tilde{\mathbf{A}} = \mathbf{D} + \mathbf{A}.
\end{align}
\noindent Here, the matrix $\mathbf{D}$ is the degree matrix of $\mathbf{A}$. Note that because the self-loop is a sum of all the weights to a node's neighbors, the initial importance of the node itself can be disproportionately large.

\textbf{$\bullet$ Symmetric Normalized Matrix w/ Row-wise Max: } Finally, we calculate a symmetric normalized matrix (similar to the normalized Laplacian) via:
\begin{align}
\label{eq:k5}
\tilde{\mathbf{A}} = \mathbf{D}^{-\frac{1}{2}} (\mathbf{A} + \mathbf{M}) \mathbf{D}^{-\frac{1}{2}}.
\end{align}
\noindent Here, based on the context, the matrix $\mathbf{D}$ is the diagonal degree matrix of $\mathbf{A} + \mathbf{M}$.

\subsubsection{K-Step Motif Matrices}
Given a step-size $K$, we further define $K$ different $k$-step motif-based matrices for each of the $T$ motifs which gives a total of $K \times T$ adjacency matrices. Formally, this is formulated as follows:
\begin{align}
\label{eq:kstep1}
\tilde{\mathbf{A}}^{(k)}_t = \Psi(\mathbf{A}^k_t), \,\, \text{for } k = 1, \cdots, K \text{ and } t = 1, \cdots, T 
\end{align}
\noindent where
\begin{align}
\label{eq:kstep2}
\Psi(\mathbf{A}^k_t) = \Psi(\underbrace{\mathbf{A}_t \cdots \mathbf{A}_t}_{k})
\end{align}

When we set $K > 1$, we allow nodes to accumulate information from a wider neighborhood. For instance if we choose to use Eq.~\ref{eq:k1} (for $\Psi$) and use an edge as our motif, $\tilde{\mathbf{A}}^{(k)}$ (we omit the motif-type subscript here) then captures $k$-hop neighborhoods of each node. While, in theory, using $\tilde{\mathbf{A}}^{(k)}$ is equivalent to a $k$-layer GCN or GAT model~\cite{N-GCN} have shown that GCNs don't necessarily benefit from a wider receptive field from increasing model depth. This may be for reasons similar as to why skip-connections are needed in deep architectures since the signal starts to degrade as the model gets deeper~\cite{skip}.

As another example, we set $\Psi$ to Eq.~\ref{eq:k3}. Now for an arbitrary motif, we see that $(\tilde{\mathbf{A}}^{(k)})_{i,j}$ encodes the probability of transitioning from node $i$ to node $j$ in $k$ steps.

\subsubsection{Motif Matrix Selection via Attention} Given $T$ different motifs and a step-size of $K$, we now have $K \times T$ motif matrices we could use with Eq.~\ref{eq:gcn} to define layer-wise propagations. A simple approach would be to implement $K \times T$ idependent GCN instances and concatenate the final node outputs before classification. However, this approach may have problems scaling when $T$ and/or $K$ is large.

Instead, we propose to use an attention mechanism, at each layer, to allow each node to select \textit{a single} most relevant neighborhood to integrate or accumulate information from. For a layer $l$, this can be defined by two functions $f_l : \mathbb{R}^{S_l} \rightarrow \mathbb{R}^{T}$ and $f_l^\prime : \mathbb{R}^{S_l}\times \mathbb{R}^{T} \rightarrow \mathbb{R}^{K}$, where $S_l$ is the dimension of the state-space for layer $l$. The functions' outputs are \textit{softmaxed} to form probability distributions over $\{1, \cdots, T\}$ and $\{1, \cdots, K\}$, respectively. Essentially, what this means is that given a node $i$'s state, the functions recommend the most relevant motif $t$ and step size $k$ for node $i$ to integrate information from.

Specifically, we define the state matrix encoding node states at layer $l$ as a concatenation of two matrices:
\begin{align}
\label{eq:state}
\mathbf{S}_l = \left[ \Psi(\mathbf{A}) \mathbf{H}^{(l)} \mathbf{W}^{(l)} \,\,\,\,\, \mathbf{C} \right],
\end{align}
\noindent where $\mathbf{W}^{(l)} \in \mathbb{R}^{N \times D_l}$ is the weight matrix that embeds the inputs to dimension $D_l$, $\Psi(\mathbf{A}) \mathbf{H}^{(l)} \mathbf{W}^{(l)}$ is the matrix containing local information obtained by doing a weighted sum of the features in the simple one-hop neighborhood for each node (from the original adjacency $\mathbf{A}$), and $\mathbf{C} \in \mathbb{R}^{N \times C}$ is a motif count matrix that gives us basic local structural information about each node by counting the number of $C$ different motifs that each node belongs to. We note here that $\mathbf{C}$ is \textit{not} appended to the node attribute matrix $\mathbf{X}$ and is not used for prediction. It's only purpose is to capture the local structural information of each node. $\mathbf{C}$ is precomputed once.

Let us consider an arbitrary layer. Recall that $f$ (for brevity, we omit subscripts $l$) produces a probability vector specifying the importance of the various motifs, let $\vec{\mathbf{f}}_i = f(\vec{\mathbf{s}}_i)$ be the motif probabilities for node $i$. Similarly, let $\vec{\mathbf{f}}_i^{\prime} = f^\prime(\vec{\mathbf{s}}_i,\vec{\mathbf{f}}_i)$ be the probability vector recommending the step size. Now let $t_i$ be the index of the largest value in $\vec{\mathbf{f}}_i$ and similarly, let $k_i$ be the index of the largest value in $\vec{\mathbf{f}}_i^{\prime}$. In other words, $t_i$ is the recommended motif for $i$ while $k_i$ is the recommended step-size. Attention can now be used to define an $N \times N$ propagation matrix as follows:
\begin{equation}
\label{eq:att_matrix}
\hat{\mathbf{A}} = \begin{bmatrix} 
    \left( \tilde{\mathbf{A}}^{(k_1)}_{t_1} \right)_{1,:} \\
    \vdots  \\
    \left( \tilde{\mathbf{A}}^{(k_N)}_{t_N} \right)_{N,:}
    \end{bmatrix}.
\end{equation}
This layer-specific matrix $\hat{\mathbf{A}}$ can now be plugged into Eq.~\ref{eq:gcn} to replace $\tilde{\mathbf{A}}$. What this does is it gives each node the flexibility to select the most appropriate motif $t$ and step-size $k$ to integrate information from. 

\subsubsection{Training the Attention Mechanism} 
Given a labeled graph $\mathcal{G} = (\mathcal{V}, \mathcal{E}, \ell)$ with $N$ nodes and a labeling function $\ell: \mathcal{V} \rightarrow \mathcal{L}$ which maps each node to one of $L$ class labels in $\mathcal{L} = \{1, \cdots, L \}$, our goal is to train a classifier that can predict the label of all the nodes. Given a subset $\mathcal{T} \subset \mathcal{V}$, or the training set of nodes, we can train an $L$-layer MCN (the classifier) using standard cross-entropy loss as follows:
\begin{align}
\label{eq:xent_loss}
\mathcal{L}_C = - \sum_{v \in \mathcal{T}}\sum_{l = 1}^{L} Y_{vl} \, \text{log}\,\pi(H^{(L+1)}_{i,l}),
\end{align}
\noindent where $Y_{vl}$ is a binary value indicating node $v$'s true label (\textit{i.e.}, $Y_{vl} = 1$ if $\ell(v) = l$, zero otherwise), and $\mathbf{H}^{L+1} \in \mathbb{R}^{N \times L}$ is the \textit{softmaxed} output of the MCN's last layer.

While Eq.~\ref{eq:xent_loss} is sufficient for training the MCN to classify inputs it does not tell us how we can train the attention mechanism that selects the best motif and step-size for each node at each layer. We define a second loss function based on reinforcement learning as follows:
\begin{align}
\label{eq:att_loss}
\resizebox{1\hsize}{!}{
$\mathcal{L}_A = - \sum_{n_L \in \mathcal{T}} R_v \left[ \text{log}\,\pi \left(\left( \vec{\mathbf{f}}^{(L)}_{n_L} \right)_{t^{(L)}_{n_L}} \right) + \text{log}\,\pi \left(\left( \vec{\mathbf{f}}^{(L)}_{n_L} \right)_{k^{(L)}_{n_L}} \right) \right] \nonumber$} \\ 
\resizebox{1\hsize}{!}{
$+ \sum_{n_L \in \mathcal{T}}\sum_{n_{L-1} \in \mathcal{N}^{(\hat{\mathbf{A}}^{(L)})}_{n_L}} R_v \left[ \text{log}\,\pi \left(\left( \vec{\mathbf{f}}^{(L-1)}_{n_{L-1}} \right)_{t^{(L-1)}_{n_{L-1}}} \right) + \text{log}\,\pi \left(\left( \vec{\mathbf{f}}^{(L-1)}_{n_{L-1}} \right)_{k^{(L-1)}_{n_{L-1}}} \right) \right] \nonumber$} \\
\resizebox{1\hsize}{!}{
$+ \cdots + \sum_{n_L \in \mathcal{T}} \cdots \sum_{n_1 \in \mathcal{N}^{(\hat{\mathbf{A}}^{(2)})}_{n_2}} R_v \left[ \text{log}\,\pi \left(\left( \vec{\mathbf{f}}^{(1)}_{n_1} \right)_{t^{(1)}_{n_1}} \right) + \text{log}\,\pi \left(\left( \vec{\mathbf{f}}^{(1)}_{n_1} \right)_{k^{(1)}_{n_1}} \right) \right]$}
\end{align}

\noindent Here, $R_v$ is the reward we give to the system ($R_v = 1$ if we classify $v$ correctly, $R_v = -1$ otherwise). The intuition here is this: at the last layer we reward the actions of the classified nodes; we then go to the previous layer (if there is one) and reward the actions of the neighbors of the classified nodes since their actions affect the outcome, we continue this process until we reach the first layer.

There are a few important things to point out. In practice, we use an $\epsilon$-greedy strategy when selecting a motif and step-size during training. Specifically, we pick the action with highest probability most of the time but during $1 - \epsilon$ instances we select a random action. During testing, we choose the action with highest probability. Also, in practice, we use dropout to train the network as in GAT~\cite{GAT} which is a good regularization technique but also has the added advantage of being a way to sample the neighborhood during training to keep the receptive field from growing too large during training. Finally, to reduce model variance we can also include an advantage term (see Eq. 2 in~\cite{GAM}, for instance). Our final loss can then be written as:
\begin{align}
\label{eq:loss}
\mathcal{L} = \mathcal{L}_C + \mathcal{L}_A.
\end{align}

\begin{figure}[t]
\centering
\includegraphics[width=1\linewidth]{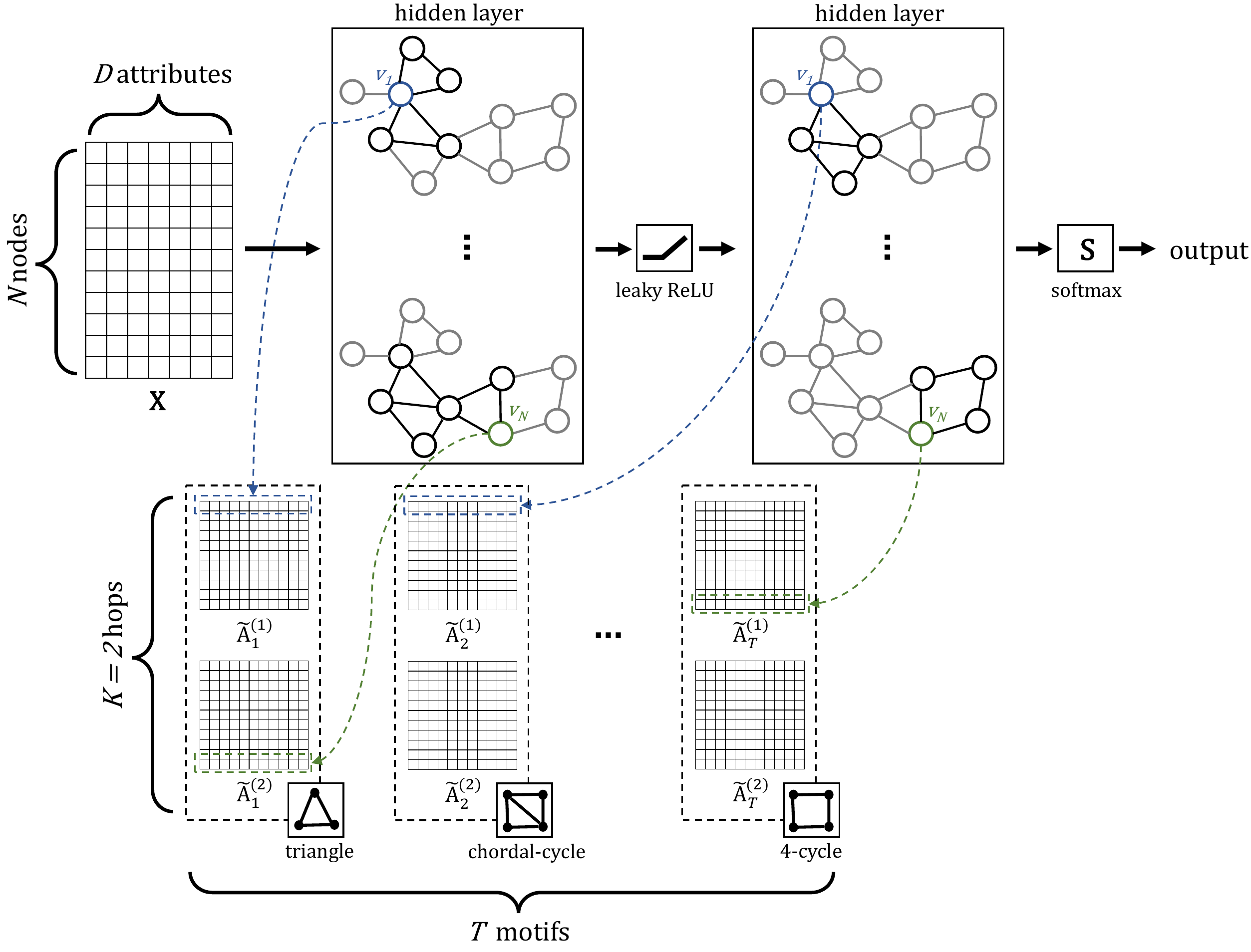}
\vspace{-7mm}
\caption{An example of a 2-layer MCN with $N = 11$ nodes, step-size $K = 2$, and $T$ motifs. Attention allows each node to select a different motif-induced neighborhood to accumulate information from for each layer. For instance, in layer $1$, the node $v_N$ considers neighbors (up to 2-hops) that share a stronger bond (in this case, triangles) with it.}
\label{fig:model}
\vspace{-5mm}
\end{figure}

We show a simple ($2$-layer) example of the proposed MCN model in Fig.~\ref{fig:model}. As mentioned, MCN generalizes both GCN and GAT. We list settings of these methods in Tab.~\ref{tab:gen}.

\begin{table}[h!]
    \centering
    \caption{Space of methods expressed by MCN. 
    GCN and GAT are shown below to be special cases of MCN.}
    \label{tab:gen}
    \vspace{-2mm}
    \resizebox{\columnwidth}{!}{
    \begin{tabular}{lccccc}
    \toprule
    &&&&&\vspace{-8pt}\\
        \textbf{Method}
        & \textbf{Motif}
        & \textbf{Adj.}
        & \textbf{K}
        & \textbf{Self-attention}
        & \textbf{Motif-attention}\\
    \midrule
    \vspace{-8pt}&&&&&\\
    {GCN} & edge & \multicolumn{1}{c}{Eq.~\ref{eq:k1}} &$K=1$&no&no\\ \vspace{-8pt}&&&&&\\
    {GAT}   & edge & Eq.~\ref{eq:k1} & $K=1$&yes&no\\ \vspace{-8pt}&&&&&\\
    {MCN-*} & any & Eqs.~\ref{eq:k1}-\ref{eq:k5}& $K=\{1,\cdots\}$& yes&yes\\
    \bottomrule
    \end{tabular}
    }
    \vspace{-5mm}
\end{table}

\begin{table*}[t]
\centering
\small
\fontsize{9.8}{11}\selectfont
\caption{Summary of experimental results: ``average accuracy $\pm$ SD ({\color{blue} rank})". The ``avg. rank" column shows the average rank of each method. The lower the average rank, the better the overall performance of the method.}
\vspace{-4mm}
\label{tab:acc}
\begin{center}
\begin{tabular}{l l l l l l l l l l@{}}
\toprule
\multicolumn{1}{c}{\multirow{2}{*}{\bf method}}  & \multicolumn{3}{c}{\rule{0pt}{2ex} \bf dataset} & \multicolumn{1}{c}{\multirow{2}{*}{\bf \shortstack{avg.\\rank}}}\\
\cline{2-4}
& \multicolumn{1}{c}{\rule{0pt}{2ex} Cora} & \multicolumn{1}{c}{Citeseer} & \multicolumn{1}{c}{Pubmed}  & \\
\midrule
{\rule{0pt}{2ex}\textbf{DeepWalk}~\cite{deepwalk}} & 67.2\% ({\color{blue} 9}) & 43.2\% ({\color{blue} 11}) & 65.3\% ({\color{blue} 11}) & \multicolumn{1}{c}{\color{blue} 10.3}\\
\textbf{MLP} & 55.1\% ({\color{blue} 12}) & 46.5\% ({\color{blue} 9}) & 71.4\% ({\color{blue} 9}) & \multicolumn{1}{c}{\color{blue} 10.0}\\
\textbf{LP}~\cite{lp} & 68.0\% ({\color{blue} 8}) & 45.3\% ({\color{blue} 10}) & 63.0\% ({\color{blue} 12}) & \multicolumn{1}{c}{\color{blue} 10.0}\\
\textbf{ManiReg}~\cite{manireg} & 59.5\% ({\color{blue} 10}) & 60.1\% ({\color{blue} 7}) & 70.7\% ({\color{blue} 10}) & \multicolumn{1}{c}{\color{blue} 9.0}\\
\textbf{SemiEmb}~\cite{semiemb} & 59.0\% ({\color{blue} 11}) & 59.6\% ({\color{blue} 8}) & 71.7\% ({\color{blue} 8}) & \multicolumn{1}{c}{\color{blue} 9.0}\\
\textbf{ICA}~\cite{ICA} & 75.1\% ({\color{blue} 7}) & 69.1\% ({\color{blue} 5}) & 73.9\% ({\color{blue} 7}) & \multicolumn{1}{c}{\color{blue} 6.3}\\
\textbf{Planetoid}~\cite{planetoid} & 75.7\% ({\color{blue} 6}) & 64.7\% ({\color{blue} 6}) & 77.2\% ({\color{blue} 5}) & \multicolumn{1}{c}{\color{blue} 5.7}\\
\textbf{Chebyshev}~\cite{cheby} & 81.2\% ({\color{blue} 5}) & 69.8\% ({\color{blue} 4}) & 74.4\% ({\color{blue} 6}) & \multicolumn{1}{c}{\color{blue} 5.0}\\
\textbf{MoNet}~\cite{monet} & 81.7\% ({\color{blue} 3}) & -- & 78.8\% ({\color{blue} 4}) & \multicolumn{1}{c}{\color{blue} 3.5}\\
\textbf{GCN}~\cite{GCN} & 81.5\% ({\color{blue} 4}) & 70.3\% ({\color{blue} 3}) & 79.0\% ({\color{blue} 2}) & \multicolumn{1}{c}{\color{blue} 3.0}\\
\textbf{GAT}~\cite{GAT} & 83.0 $\pm$ 0.7\% ({\color{blue} 2}) & 72.5 $\pm$ 0.7\% ({\color{blue} 2}) & 79.0 $\pm$ 0.3\% ({\color{blue} 2}) & \multicolumn{1}{c}{\color{blue} 2.0}\\

\textbf{MCN} (this paper) & 83.5 $\pm$ 0.4\% ({\color{blue} 1}) & 73.3 $\pm$ 0.7\% ({\color{blue} 1}) & 79.3 $\pm$ 0.3\% ({\color{blue} 1}) & \multicolumn{1}{c}{\color{blue} 1.0}\\
\bottomrule
\end{tabular}
\vspace{-5mm}
\end{center}
\end{table*}

\section{Experimental Results}
\subsection{Semi-supervised node classification} We first compare our proposed approach against a set of strong baselines (including methods that are considered the current state-of-the-art) on three well-known graph benchmark datasets for semi-supervised node classification. We show that the proposed method is able to achieve state-of-the-art results on all compared datasets.

\subsubsection{Datasets} The datasets used for comparison are Cora, Citeseer, and Pubmed. Specifically, we use the pre-processed version made available by~\citet{planetoid}. The aforementioned graphs are undirected citation networks where nodes represent documents and edges denote citation; furthermore, a bag-of-words vector capturing word counts in each document serves as each node's feature. Each document is assigned a class label. 

The graph in Cora has $|\mathcal{V}| = 2,708$, $|\mathcal{E}| = 5,429$, 7 classes, and $D = 1,433$ node features. The statistics for Citeseer are $|\mathcal{V}| = 3,327$, $|\mathcal{E}| = 4,732$, with 6 classes, and $D = 3,703$. Finally, Pubmed consists of a graph with $|\mathcal{V}| = 19,717$, $|\mathcal{E}| = 44,338$, with 3 classes, and $D = 500$. Following previous work, we use \textit{only} 20 nodes per class for training~\cite{planetoid,GCN,GAT}. Again, following the procedure in previous work, we take 1,000 nodes per dataset for testing and further take an additional 500 for validation as in~\cite{GCN,GAT,N-GCN}. We use the same train/test/validation splits as defined in~\cite{GCN,GAT}.

\subsubsection{Setup} For Cora and Citeseer, we used the same $2$-layer model architecture as that of GAT consisting of $8$ self-attention heads each with a total of $8$ hidden nodes (for a total of 64 hidden nodes) in the first layer, followed by a single softmax layer for classification~\cite{GAT}. Similarly, we fixed early-stopping patience at $100$ and $\ell_2$-regularization at $0.0005$. For Pubmed, we also used the same architecture as that of GAT (first layer remains the same but the output layer has $8$ attention heads to deal with sparsity in the training data). Patience remains the same and similar to GAT, we use a strong $\ell_2$-regularization at $0.001$.

We further optimized all models by testing dropout values of $\{0.50, 0.55, 0.60, 0.65\}$, learning rates of $\{0.05, 0.005\}$, step-sizes $K \in \{1, 2, 3\}$, and motif adjacencies formed using combinations of the following motifs: \textit{edge}, \textit{2-star}, \textit{triangle}, \textit{3-star}, and \textit{4-clique} (please refer to Tab.~\ref{fig:butterfly}(d) for motifs). 
Self-attention learns to prioritize neighboring features that are more relevant. $\Psi$ (Eqs.~\ref{eq:k1}-\ref{eq:k5}) can be used as a reasonable initial estimate of the importance of neighboring features. For each unique setting of the hyperparameters mentioned previously, we try Eqs.~\ref{eq:k1}-\ref{eq:k5} and record the best result. Finally, we adopt an $\epsilon$-greedy strategy ($\epsilon=0.1$).

\subsubsection{Comparison} For all three datasets, we report the classification accuracy averaged over $15$ runs on random seeds (including standard deviation). Since we utilize the same train/test splits as previous work, we follow~\cite{GAT,GCN} and compile all previously reported results.

A summary of the results is shown in Tab.~\ref{tab:acc}. We see that our proposed method achieves superior performance against all tested baselines on all three benchmarks. On the Cora dataset, the best model used a learning rate of $0.005$, dropout of $0.6$, and both the edge and triangle motifs with step-size $K=1$. For Citeseer, the learning rate was $0.05$ and dropout was still $0.6$ while the only motif used was the edge motif with step-size $K=2$. However, the second best model for Citeseer -- which had comparable performance -- utilized the following motifs: edge, 2-star, and triangle. Finally, on Pubmed, the best model used learning rate $0.05$ and dropout of $0.5$. Once again, the best motifs were the edge and triangle motifs on $K=1$.

We find that the triangle motif is useful in improving classification performance on the compared datasets. This highlights an advantage of MCN over past approaches (\textit{e.g.}, GCN \& GAT) which do not handle triangles (and other motifs) naturally. The results seem to indicate that it can be beneficial to consider stronger bonds (friends that are friends themselves) when selecting a neighborhood.

\subsection{Comparison on Datasets exhibiting Heterophily} The benchmark datasets (Cora, Citeseer, and Pubmed) that we initially tested our method on exhibited strong homophily where nodes that share the same labels tend to form densely connected communities. Under these circumstances, methods like GAT or GCN that use a first-order propagation rule will perform reasonably well. However, not all real-world graphs share this characteristic and in some cases the node labels are more spread out. In this latter case, there is reason to believe that neighborhoods constructed using different motifs -- other than just edges and triangles -- may be beneficial. 

We test this hypothesis by comparing GAT and GCN against MCN on two graphs from the DD dataset~\cite{KKMMN2016}. Specifically, we chose two of the largest graphs in the dataset: DD-6 and DD-7 -- with a total of $4,152$ and $1,396$ nodes, respectively. Both graphs had twenty different node labels with the labels being quite imbalanced. 

We stick to the semi-supervised training regime, using only $15$ nodes per class for training with the rest of the nodes split evenly between testing and validation. This makes the problem highly challenging since the graphs do not exhibit homophily. Since the nodes do not have any attributes, we use the well-known Weisfeiler-Lehman algorithm (we initialize node attributes to a single value and run the algorithm for 3 iterations) to generate node attributes that capture each node's neighborhood structure.

For the three approaches (GCN, GAT, and MCN), we fix early-stop patience at $50$ and use a two-layer architecture with 32 hidden nodes in the first layer followed by the softmax output. We optimized the hyperparameters by searching over learning rate in $\{0.05, 0.005\}$, $\ell_2$ regularization in $\{0.01, 0.001, 0.0001, 0.00001$, dropout at $\{ 0.2, 0.3, 0.4, 0.5, 0.6\}$. Furthermore, for MCN, we considered combinations of the following motifs \{edge, 2-star, triangle, 4-path-edge, 3-star, 4-cycle, 4-clique\} and considered $K$-steps from $1, \cdots, 4$. Since there are multiple classes and they are highly imbalanced, we report the Micro-F1 score averaged over 10 runs. 

\begin{table}[t]
\caption{Micro-F1 scores of compared methods on DD.}
\vspace{-2mm}
\label{tab:DD}
\begin{center}
\begin{tabular}{l l l l l l l l }
\toprule
\multicolumn{1}{c}{\multirow{2}{*}{\bf method}}  & \multicolumn{2}{c}{\rule{0pt}{2ex} \bf dataset}\\
\cline{2-3}
& \multicolumn{1}{c}{\rule{0pt}{2ex} DD-6} & \multicolumn{1}{c}{DD-7} \\
\hline
{\rule{0pt}{2ex}GCN} & $11.9 \pm 0.6$\% & $12.4 \pm 0.8$\% \\
GAT & $11.8 \pm 0.5$\% & $11.8 \pm 1.1$\%  \\
MCN & $\mathbf{12.4 \pm 0.5}\%$ & $\mathbf{13.1 \pm 0.9\%}$ \\

\bottomrule
\end{tabular}
\vspace{-5mm}
\end{center}
\end{table}

A summary of the results is shown in Tab.~\ref{tab:DD}. While the methods do not perform very well (to be expected since we use a very small subset for training on graphs that do not have a high degree of homophily) we do find that with everything else constant (model architecture), it is actually valuable to use motifs. For DD-6, the best method utilized all motifs except for the 4-path-edge with $K=1$ while in DD-7 the best approach only used the edge, triangle, and 4-clique motifs with $K=4$. 

\begin{figure}[t]
\centering
\includegraphics[width=1.0\linewidth]{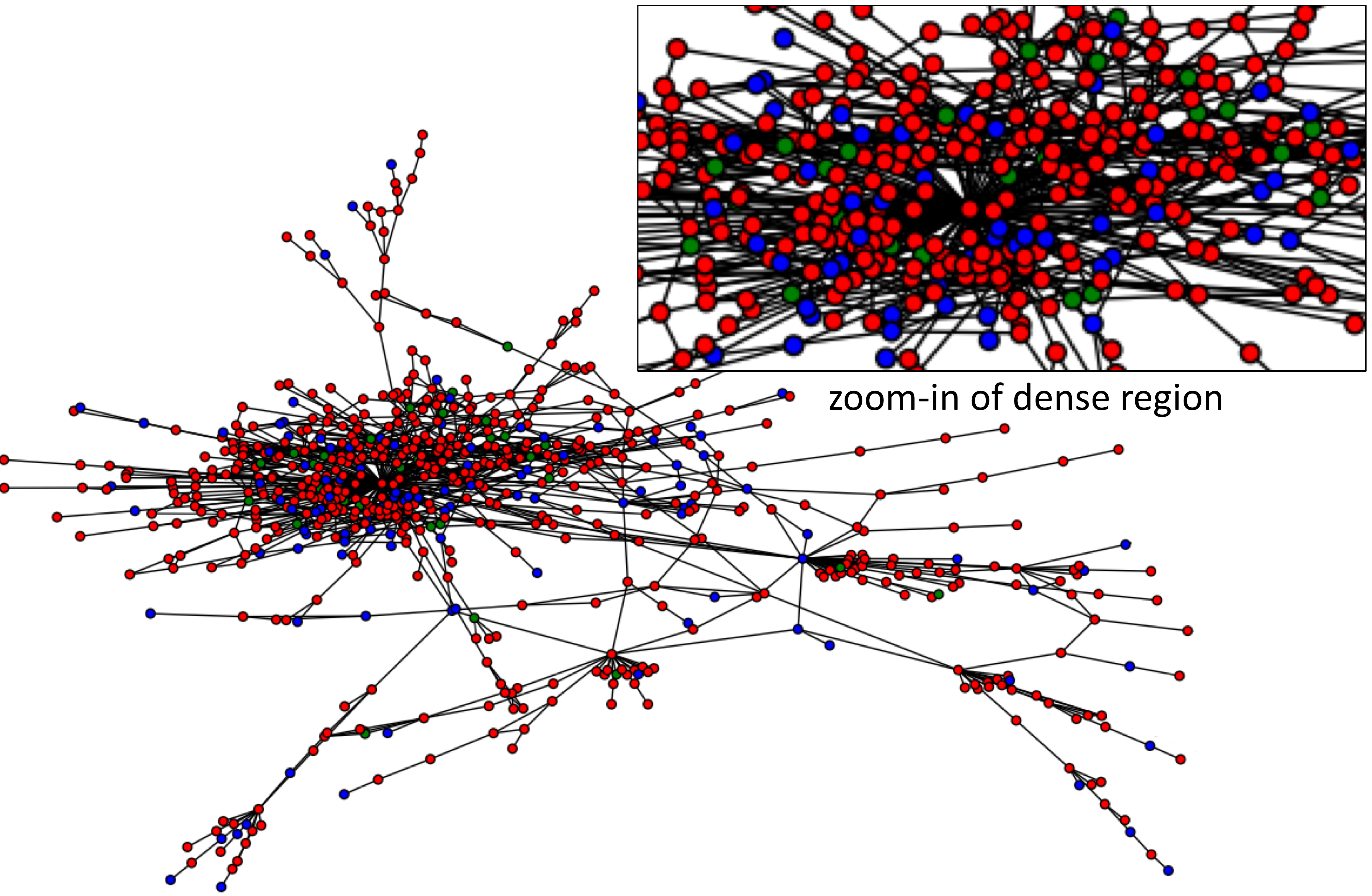}
\vspace{-7mm}
\caption{Nodes in the largest connected component of an induced subgraph in Cora (containing nodes from class $4$). The nodes are colored to indicate the motif chosen by the attention mechanism in the first layer. The motifs are: edge ({\color{blue}blue}), 4-path ({\color{red}red}), and triangle ({\color{green}green}). We observe that the nodes near the fringe of the network tend to select the 4-path allowing them to aggregate information from a wider neighborhood. On the other hand, nodes that choose triangle motifs can be found in the denser regions where it may be helpful to take advantage of stronger bonds.}
\label{fig:vis}
\vspace{-5mm}
\end{figure}

\subsection{Visualizing Motif Attention}
We ran an instance of MCN with the following motifs: edge, 4-path, and triangle with $K=1$ on the Cora dataset. Fig.~\ref{fig:vis} shows the motif that was selected by the attention mechanism. A few interesting things can be observed here. First, the nodes at the fringe prioritized the 4-path motif which is quite intuitive since this allows the nodes to aggregate information from a wider (4-hop) neighborhood which is useful for nodes near the fringe that are more separated from the other nodes in the class. On the other hand, we observe that nodes that chose the triangle motif are almost always found in denser parts of the graph. This shows that it may be beneficial in these cases to consider stronger bonds (\textit{e.g.}, when the neighborhood is noisy). Finally, we see that attention allows different nodes to select different motifs and the system is not ``defaulting" to a single motif type.

\section{Conclusion}
In this work, we introduce the Motif Convolutional Network which uses a novel attention mechanism to allow different nodes to select a different neighborhood to integrate information from. The method generalizes both GAT and GCN. Experiments on three citation (Cora, Citeseer, \& Pubmed) and two bioinformatic (DD-6 \& DD-7) benchmark graphs show the advantage of the proposed approach. We also show experimentally that different nodes do utilize attention to select different neighborhoods, indicating that it may be useful to consider various motif-defined neighborhoods.

\fontsize{9.0pt}{10.0pt}
\bibliography{paper}
\bibliographystyle{aaai}
\end{document}